\documentclass[12pt,a4paper]{article}
\usepackage{amsmath,amssymb,url}
\usepackage{hyperref}
\hypersetup{
    colorlinks = true,
    citecolor = {blue}
}
\usepackage{graphicx,tabularx}
\usepackage{array,dsfont}

\makeatletter
\newif\if@preliminary
\@preliminaryfalse
\def\preliminary{\@preliminarytrue}
\def\preprintno#1{\def\@preprintno{#1}}
\def\address#1{\def\@address{#1}}
\def\email#1#2{\thanks{\tt #1@{}#2}}
\def\abstract#1{\def\@abstract{#1}}
\def\talk#1{\def\@talk{#1}}
\renewcommand\abstractname{ABSTRACT}

\newlength\preprintnoskip
\newlength\abstractwidth
\newlength\talkwidth
\@titlepagetrue
\renewcommand\maketitle{\begin{titlepage}%
  \let\footnotesize\small
  \hfill\parbox{\preprintnoskip}{%
  \begin{flushright}\@preprintno\end{flushright}}\hspace*{1cm}
  \vskip 40\p@
  \begin{center}%
    {\Large\bf\boldmath \@title \par}\vskip 1cm%
    {\sc\@author \par}\vskip 3mm%
    {\@address \par}%
    \if@preliminary
      \vskip 2cm {\large\sf PRELIMINARY DRAFT \par \@date}%
    \fi
  \end{center}\par
  \vskip 4mm
  \begin{center}%
    \parbox{\talkwidth}{
      \begin{center}
        {\em Talk presented by J\"urgen Reuter at the International
          Workshop on Future Linear Colliders (LCWS2017), Strasbourg,
          France, 23-27 October 2017. C17-10-23.2.}
      \end{center}
    }
  \end{center}
  \vskip .4cm
  \begin{center}%
    \parbox{\abstractwidth}{\centerline{\abstractname}%
    \vskip 3mm%
    \@abstract}
  \end{center}
  \vfill
  \@thanks    
  \end{titlepage}%
  \setcounter{footnote}{0}%
  \let\thanks\relax\let\maketitle\relax
  \gdef\@thanks{}\gdef\@author{}\gdef\@address{}%
  \gdef\@title{}\gdef\@abstract{}\gdef\@preprintno{}
}%
\def\@citex[#1]#2{\if@filesw\immediate\write\@auxout{\string\citation{#2}}\fi
  \def\@citea{}\@cite{\@for\@citeb:=#2\do
    {\@citea\def\@citea{,\penalty\@m}\@ifundefined
       {b@\@citeb}{{\bf ?}\@warning
       {Citation `\@citeb' on page \thepage \space undefined}}%
\hbox{\csname b@\@citeb\endcsname}}}{#1}}
\def\citerange{\@ifnextchar [{\@tempswatrue\@citexr}{\@tempswafalse\@citexr[]}}
\def\@citexr[#1]#2{\if@filesw\immediate\write\@auxout{\string\citation{#2}}\fi
  \def\@citea{}\@cite{\@for\@citeb:=#2\do
    {\@citea\def\@citea{--\penalty\@m}\@ifundefined
       {b@\@citeb}{{\bf ?}\@warning
       {Citation `\@citeb' on page \thepage \space undefined}}%
\hbox{\csname b@\@citeb\endcsname}}}{#1}}
\long\def\@makecaption#1#2{%
  \sbox\@tempboxa{#1: \emph{#2}}%
  \ifdim \wd\@tempboxa >\hsize
    #1: \emph{#2}\par
  \else
    \hbox to\hsize{\hfil\box\@tempboxa\hfil}%
  \fi
  \vskip\belowcaptionskip}
\def\fmslash{\@ifnextchar[{\fmsl@sh}{\fmsl@sh[0mu]}}
\def\fmsl@sh[#1]#2{%
  \mathchoice
    {\@fmsl@sh\displaystyle{#1}{#2}}%
    {\@fmsl@sh\textstyle{#1}{#2}}%
    {\@fmsl@sh\scriptstyle{#1}{#2}}%
    {\@fmsl@sh\scriptscriptstyle{#1}{#2}}}
\def\@fmsl@sh#1#2#3{\m@th\ooalign{$\hfil#1\mkern#2/\hfil$\crcr$#1#3$}}
\makeatother

\newcommand\ltap{\
  \raise.3ex\hbox{$<$\kern-.75em\lower1ex\hbox{$\sim$}}\ }
\newcommand\gtap{\
  \raise.3ex\hbox{$>$\kern-.75em\lower1ex\hbox{$\sim$}}\ }

\newcommand\simge{\mathrel{%
   \rlap{\raise 0.511ex \hbox{$>$}}{\lower 0.511ex \hbox{$\sim$}}}}
\newcommand\simle{\mathrel{
   \rlap{\raise 0.511ex \hbox{$<$}}{\lower 0.511ex \hbox{$\sim$}}}}

\newcommand\be{\begin{equation}}
\newcommand\ee{\end{equation}}
\newcommand\bea{\begin{eqnarray}}
\newcommand\eea{\end{eqnarray}}
\newcommand\ba{\begin{array}}
\newcommand\ea{\end{array}}

\newcommand\whizard{\texttt{WHIZARD}}
\newcommand\openloops{\texttt{OpenLoops}}

\def\bq{\begin{equation}}
\def\eq{\end{equation}}
\def\ba{\begin{eqnarray}}
\def\ea{\end{eqnarray}}

\begin{document}

\date{\today}

\preprintno{DESY 18-013, LTH 1151, MITP/18-007, SI-HEP-2018-07, UWThPh2018-2}

\title{Exclusive top production at a Linear Collider at and off the threshold}

\author{J\"urgen Reuter\email{juergen.reuter}{desy.de}$^a$,
  Fabian Bach\email{fabian.bach}{desy.de}$^b$,
  Bijan Chokouf\'{e} Nejad\email{bijan.chokoufe}{desy.de}$^a$,
  Andre Hoang\email{andre.hoang}{univie.ac.at}$^{c,d}$,
  Wolfgang Kilian\email{kilian}{physik.uni-siegen.de}$^e$,
  Jonas Lindert\email{jonas.m.lindert}{durham.ac.uk}$^f$,
  Stefano Pozzorini\email{pozzorin}{physik.uzh.ch}$^g$,
  Maximilian Stahlhofen\email{mastahlh}{uni-mainz.de$^h$},
  Thomas Teubner\email{thomas.teubner}{liverpool.ac.uk}$^i$,
  Christian Weiss\email{christian.weiss}{desy.de}$^a$
  }

\address{\it%
$^a$DESY Theory Group, 
  Notkestr. 85, D-22607 Hamburg, Germany
\\[.5\baselineskip]
$^b$
European Commission, Eurostat, 2920 Luxembourg
\\[.5\baselineskip]
$^c$
University of Vienna, Faculty of Physics, Bolzmanngasse 5,
A-1090 Wien, Austria 
\\[.5\baselineskip]
$^d$
Erwin Schr\"odinger International Institute for Mathematical Physics,
University of Vienna, 
Boltzmanngasse 9, A-1090 Vienna, Austria
\\[.5\baselineskip]
$^e$
University of Siegen, Department of Physics,
Walter-Flex-Str. 3, D-57068 Siegen, Germany
\\[.5\baselineskip]
$^f$
Institute for Particle Physics Phenomenology, 
Ogden Centre for Fundamental Physics,
Department of Physics,
University of Durham,
Durham DH1 3LE,
United Kingdom
\\[.5\baselineskip]
$^g$
Physik-Institut, Universit\"at Z\"urich, Winterthurerstrasse 190,
CH-8057 Z\"urich, Switzerland
\\[.5\baselineskip]
$^h$
PRISMA Cluster of Excellence, Institute of Physics, Johannes Gutenberg
University, Staudingerweg 7, D-55128 Mainz, Germany
\\[.5\baselineskip]
$^i$
University of Liverpool, Department of Mathematical Sciences,
Liverpool L69 3BX, United Kingdom
}

\abstract{
  We review exclusive top pair production including decays at a future
  high-energy lepton collider, both in the threshold region and for
  higher energies. For the continuum process, we take complete QCD
  next-to-leading order matrix elements for the $2\to 6$ process with
  leptonic $W$ decays into account. At threshold, we match the
  fixed-order relativistic QCD-NLO cross section to a nonrelativistic
  cross section with next-to-leading
  logarithmic (NLL) threshold resummation implemented via a form factor.
}

\maketitle\nopagebreak


\section{Introduction}

Top physics is one of the standard pillars of the physics program of
any future high-energy lepton collider. The rationale is to determine
the properties of the heaviest Standard Model (SM) quark, its mass,
its width and its couplings to a level of accuracy that is not
possible at hadron colliders like Tevatron and the LHC, and to use
the top quark as a handle to search for physics beyond the Standard
Model. Here, we review recent work on the precision calculation of the
QCD next-to-leading order (QCD-NLO) for off-shell top quark
production, including (leptonic) $W$-decays, in the
continuum~\cite{Nejad:2016bci}. The details of this calculation and
the results will be discussed in Sec.~\ref{sec:continuum}. As shown in
this section, the perturbative fixed-order calculation yields naive K
factors of three and more in the kinematic region close to the
top-antitop production threshold for top-pair production, or better
$WbWb$ or $\ell\nu\ell\nu bb$ production. In the threshold region,
fixed-order perturbation theory in the
strong coupling $\alpha_s$ is not a good approximation anymore, but
the top velocity $v$ is an additional expansion parameter and
Coulomb-singular terms $\sim (\alpha_s/v)^n$ and (ideally also) large
logarithms $\sim(\alpha_s \log v)^n$ have to be resummed. In recent 
work~\cite{Bach:2017ggt}, we used a previously known effective field
theory setup based on non-relativistic QCD (NRQCD) to compute a form
factor accounting for the resummation of the threshold-singular terms
at NLL accuracy, implemented it in the fixed-order calculation and
matched the result to the QCD-NLO cross section in the transition
region between threshold and continuum. We thus obtained a
fully-differential cross section, which gives reliable predictions for
all center-of-mass energies. Depending on how inclusive the process
is, we achieve LL + QCD-NLO (for very exclusive processes) or NLL +
QCD-NLO precision (for inclusive processes) in the threshold
region. This will be discussed in Sec.~\ref{sec:threshold}. Finally,
we summarize and give an outlook in Sec.~\ref{sec:summary}. 


\section{Continuum calculation}
\label{sec:continuum}

We first discuss the relativistic QCD-NLO corrections to the off-shell
top pair production in the continuum, i.e. away from the production
threshold. This means, investigating either the process $e^+e^- \to
W^+ b W^- \bar{b}$ or $e^+e^- \to \ell^+ e^- \bar{\nu}_e \mu^+\nu_\mu b
\bar{b}$ including leptonic $W$ decays. Looking at the full four- or
six-particle final state, there are double-resonant diagrams included
(involving a top and an anti-top propagator), single-resonant diagrams
and non-resonant irreducible background processes.

In order to study the QCD-NLO corrections for the top production
processes discussed here, we take the \whizard\ framework for (QCD-)NLO
processes. \whizard~\cite{Kilian:2007gr} is a multi-purpose event
generator. It comes with its own matrix-element generator for 
tree-level amplitudes,
\texttt{O'Mega}~\cite{Moretti:2001zz,Nejad:2014sqa} which supports
almost arbitrary models like e.g. supersymmetry~\cite{Ohl:2002jp} and
many more. External models can be used inside \whizard\ by its
interface to \texttt{FeynRules}~\cite{Christensen:2010wz}. For the QCD
color decomposition, \whizard\ uses the color-flow
formalism~\cite{Kilian:2012pz}, and it comes with its own parton
\begin{figure}
  \begin{center}
    \includegraphics[width=.48\textwidth]
                    {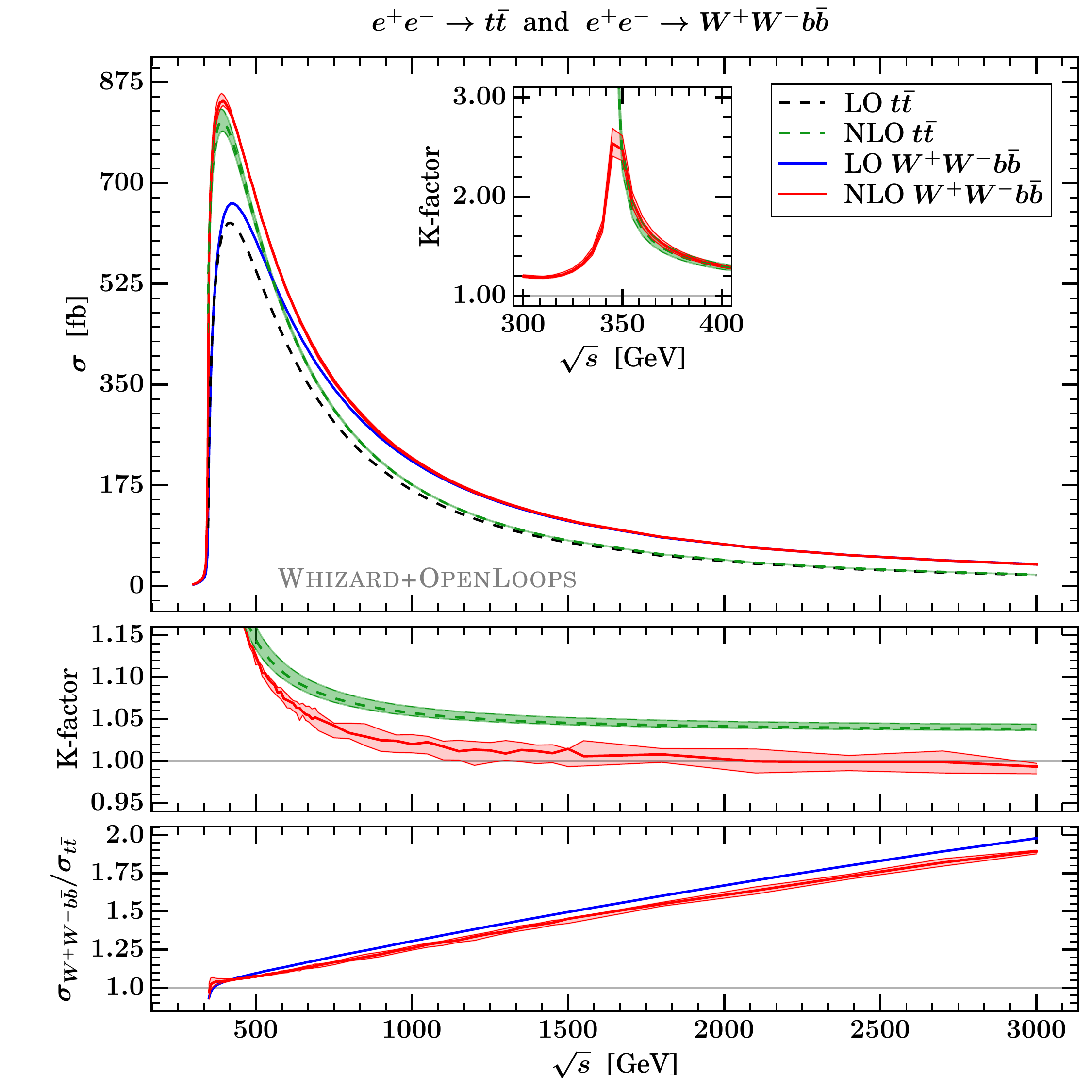}
    \includegraphics[width=.48\textwidth]
                    {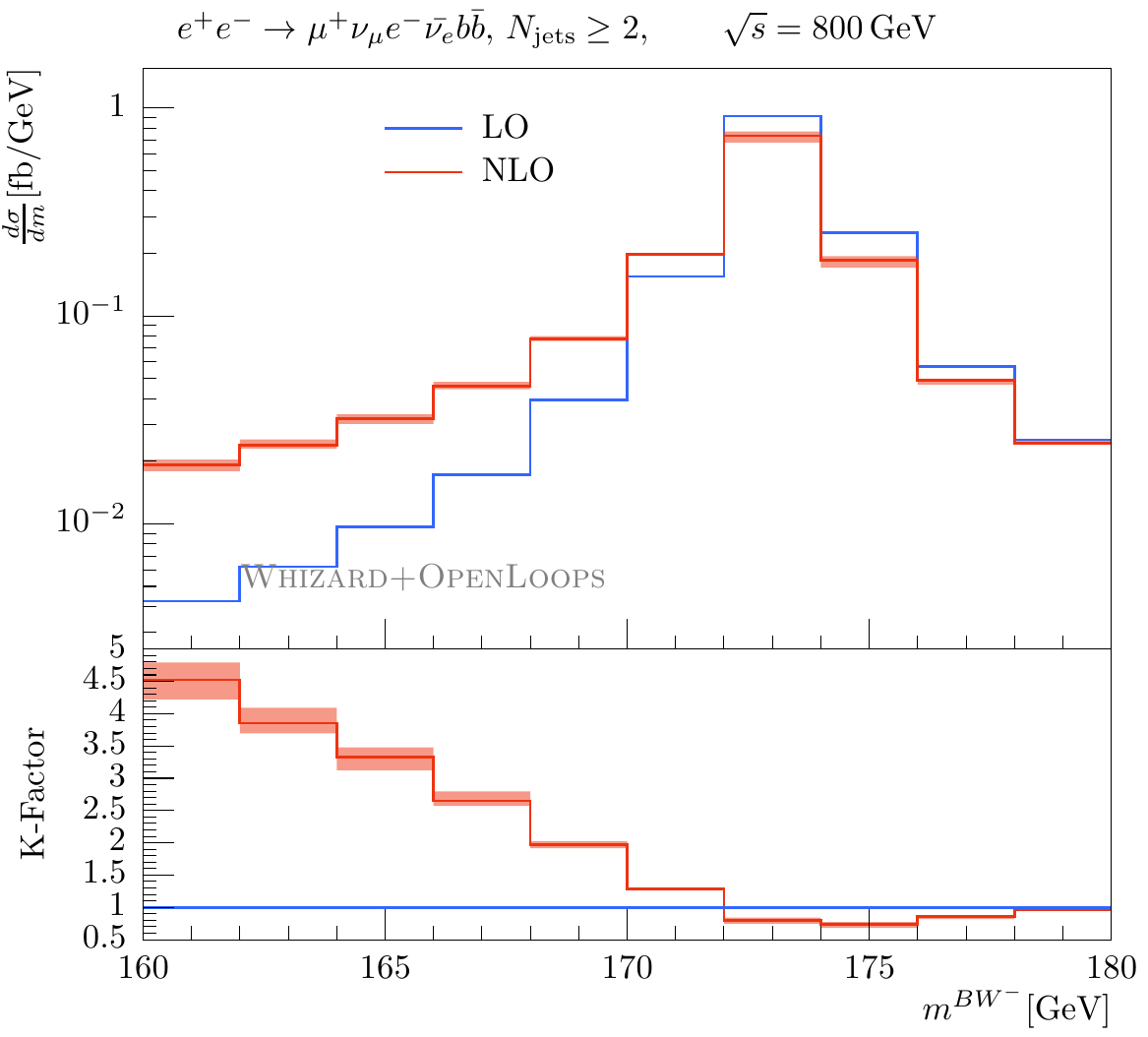}
  \end{center}
  \caption{Total cross section for the process $e^+e^- \to W^+W^- b
    \bar{b}$ as a function of the center-of-mass energy (left); the
    upper ratio plot in the bottom shows the K factor, the lower one
    the ratio of the four-body process to on-shell $t\bar{t}$
    production. The inset shows the K factor enhancement in the
    threshold region. Fully off-shell leptonic $t\bar t$ production
    ($e^+e^- \to e^-\bar{\nu}_\ell\mu^+\nu_\mu b \bar{b}$)
    differential distribution (right). Blue is LO, red is QCD-NLO
    including scale variations. Schemes and input parameters as
    described in the text.}
  \label{fig:tt_fix}  
\end{figure}
shower implementation~\cite{Kilian:2011ka}. The work on QCD-NLO within
\whizard\ started with a hard-coded implementation for the production
of $b$ jets at LHC~\cite{Binoth:2009rv,Greiner:2011mp}, while matching
between resummed terms and fixed-order calculations have been tackled
by combining fixed-order electroweak corrections to chargino
production at the ILC with an all-order QED initial-state structure
function~\cite{Kilian:2006cj,Robens:2008sa}. Since quite recently,
\whizard\ is able to do automatic POWHEG matching for $e^+e^-$
processes~\cite{Reuter:2016qbi}.

For (QCD-)NLO processes, \whizard\ automatically sets up FKS
subtraction~\cite{Frixione:1995ms} and generates the corresponding
phase space for all the singular emission regions. Here we take
virtual matrix elements, which contain up to pentagon integrals for
the processes considered here, as well as the color-correlated and
spin-correlated matrix elements for the collinear and soft splittings,
respectively, from the \openloops\ one-loop provider
(OLP)~\cite{Cascioli:2011va,Buccioni:2017yxi,Buccioni:2018zuy}. For
the amplitudes, we take the complex mass scheme using complexified
masses $\mu_i^2 = m_i^2 + i m_i \Gamma_i$, which leads e.g. to a
complex weak mixing angle. The input values are as follows: $m_W =
80.385$ GeV, $m_Z = 91.1876$ GeV, $m_t = 173.2$ GeV, $m_H = 125$ GeV,
and we perform the calculation here and for the threshold with massive
$b$-quarks of $m_b = 4.2$ GeV. It is important that the widths used in
the calculation are at the same order and in the same scheme than the
scattering process in order to guarantee properly normalized branching
ratios: $\Gamma_Z^{\text{LO}} = 2.4409$ GeV,  $\Gamma_Z^{\text{NLO}} =
2.5060$ GeV, $\Gamma_W^{\text{LO}} = 2.0454$ GeV,
$\Gamma_W^{\text{NLO}} = 2.0978$ GeV,  $\Gamma_{t\to Wb}^{\text{LO}} =
1.4986$ GeV, $\Gamma_{t\to Wb}^{\text{LO}} = 1.3681$ GeV,
$\Gamma_{t \to f\bar{f}b}^{\text{LO}} = 1.4757$ GeV,
$\Gamma_{t\to f\bar{f}b}^{\text{NLO}} = 1.3475$ GeV. Note that for the
process with stable $W$s, $e^+e^- \to W^+W^- b \bar{b}$, one has to
take the total top width for two-body decays, while for the full
process, $e^+e^- \to e^-\bar{\nu}_\ell\mu^+\nu_\nu b \bar{b}$ the
total top width for three-body decays. For the Higgs boson we take the
value $\Gamma_H = 4.31$ MeV. As the matrix elements for the full
off-shell processes contain narrow resonances, particularly the $H\to
bb$ resonance, we use a resonance-aware version of the FKS subtraction
formalism to make sure that cancellations between real emissions and
subtraction terms do cancel though the real emission could shift the
kinematics on or off the resonance compared to Born kinematics. This
resonance-aware treatment is automatically done in \whizard.
As we are using massive $b$-quarks, no cuts are necessary for the 
process $e^+e^- \to W^+W^- b \bar{b}$, but the process $e^+e^- \to
e^-\bar{\nu}_\ell\mu^+\nu_\mu b \bar{b}$ exhibits a collinear
singularity from photon emission off the electron line going from the
initial to the final state. The integrations for the full QCD-NLO
are very stable. We did two independent own integrations with
the serial and the non-blocking MPI-parallelizable version of
VAMP~\cite{Ohl:1998jn} inside \whizard, and we confirmed the result
within \texttt{Sherpa}~\cite{Gleisberg:2008ta} and \texttt{Munich}.

For the QCD-NLO corrections, we take the top mass as renormalization
scale. The scale variations for the process $e^+e^- \to W^+ b W^-
\bar{b}$ is very small, at the level of two per cent. After one has
replaced the top width in the matrix elements by a running top width
$\Gamma_t(\mu_R)$ , the scale variations for the on-shell process
$e^+e^- \to t\bar{t}$ behave the same way as for the off-shell
process. In Fig.~\ref{fig:tt_fix} we show in the left panel the total
cross section for $e^+e^- \to W^+ b W^- \bar{b}$ as a function of the
center-of-mass energy $\sqrt{s}$ over the whole kinematic range from
well below the threshold up to full energy stage of CLIC at 3 TeV.
Below the main plot there are two ratio plots, the first showing the K
factor $\sigma(NLO)/\sigma(LO)$, the second showing the ratio of the
off-shell to the on-shell process, $\sigma(e^+e^- \to W^+ b W^-
\bar{b})/\sigma(e^+e^- \to t\bar{t})$. In the upper ratio plot, the
green curve is the K factor for the on-shell process, while the red
one is the K factor for the off-shell process. For the off-shell
process, the K factor tends to almost unity at a TeV and beyond. The
second ratio plot  shows that without restricting to single- and
double resonant phase-space regions, the background processes start to
more and more dominate over the signal top-pair production. The colors
in the second ratio plot correspond to the legend, i.e. blue for LO
and red for NLO. The inset in the left plot of Fig.~\ref{fig:tt_fix}
shows the enhancement in the K factor around the top threshold where
fixed-order perturbation theory is not a valid approximation any more,
cf. Sec.~\ref{sec:threshold}. The K factor for the on-shell process
even becomes infinite here. In the right panel of
Fig.~\ref{fig:tt_fix} we show as an example for a differential
distribution the invariant mass for the negatively charged $W$
(reconstructed at Monte Carlo truth level) and the $b$ jet for the
full process $e^+e^- \to e^-\bar{\nu}_\ell\mu^+\nu_\mu b \bar{b}$ at
the center-of-mass energy of $\sqrt{s} = 800$ GeV where the cross
section peaks. The LO distribution is shown in blue and the QCD-NLO
distribution including scale variations in red. One clearly sees that
the K factor is not constant over the phase space. Particularly for
low invariant masses below the top mass peak, there is a large
enhancement as this part of the phase space gets populated massively
by gluon radiation off the top peak region. 

Note that the setup inside \whizard\ allows to immediately do QCD-NLO
calculations and simulations for polarized cross sections, or include
QED initial-state photon radiation as well as effects from
beamspectra. In Ref.~\cite{Nejad:2016bci}, we also calculated the
processes $e^+e^- \to t\bar{t}H$, $e^+e^- \to W^+ bW^- \bar{b}H$ and
$e^+e^-\to e^-\bar{\nu}_e \mu^+ \nu_\mu b\bar{b}H$ at QCD-LO and
QCD-NLO, which we omitted in this proceedings article here.

 
\section{Threshold matching}
\label{sec:threshold}

A kinematic fit to the shape of the rising of the cross section at the
top threshold is believed to be the most precise method to measure the
top quark mass with an ultimate precision of 30-80 MeV. For this the
systematic uncertainties of the experimental measurement -- especially
the details of the beam spectrum -- as well as the theoretical
uncertainties have to be well under control. As shown
Sec.~\ref{sec:continuum}, close to the kinematical threshold 
for the on-shell production of a $t\bar{t}$ 
\begin{figure}
  \begin{center}
    \includegraphics[width=.48\textwidth]
                    {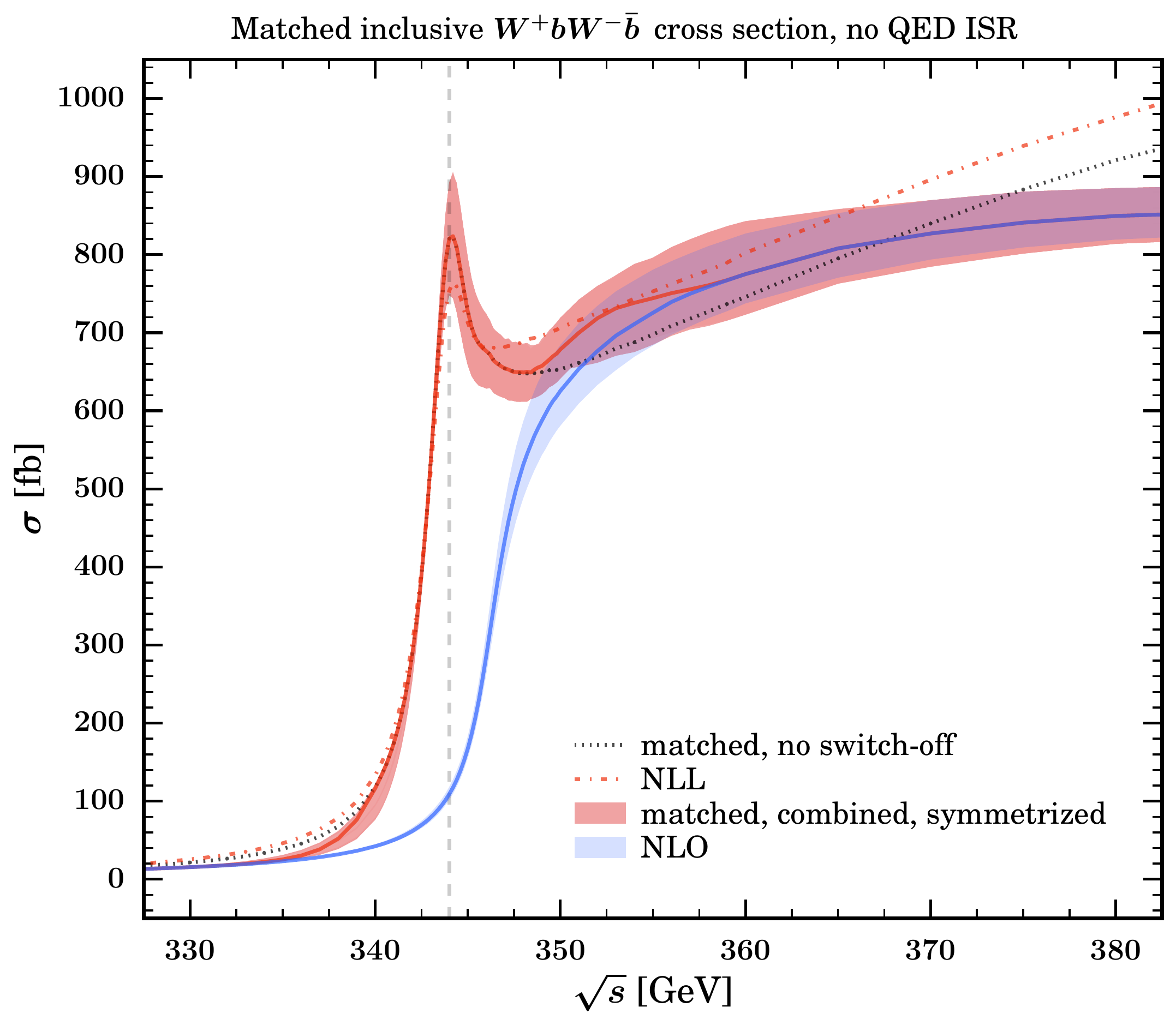}
    \includegraphics[width=.48\textwidth]
                    {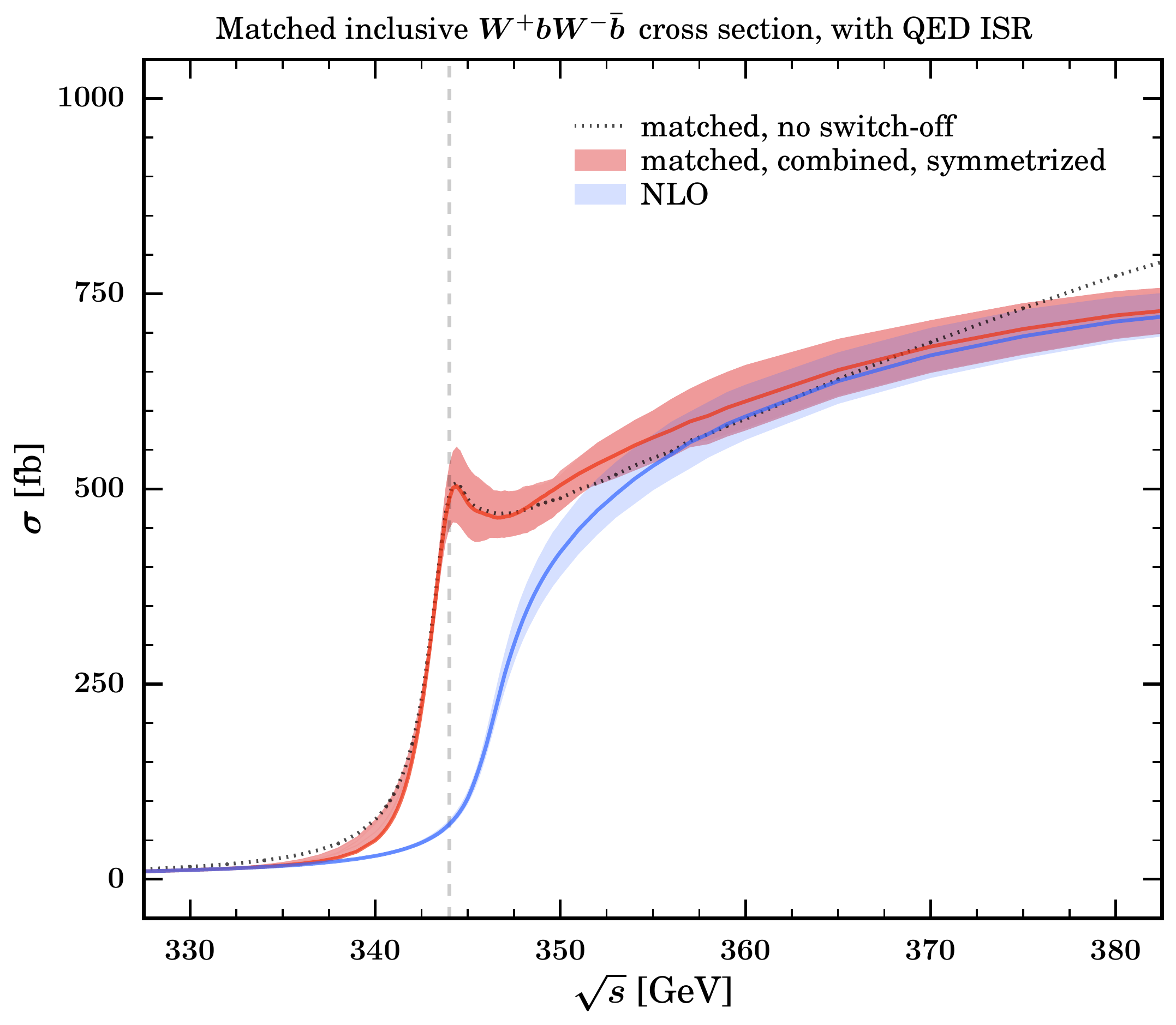}    
  \end{center}
  \caption{Matched NRQCD-NLL + QCD-NLO 
    calculation without (left) and with (right) QED ISR. The dashed
    vertical line is the value of twice $M^{1\text{S}}$. Blue is the
    fixed QCD-NLO calculation, red is the fully matched
    calculation. The matched calculation has a full envelope over
    (symmetrized) scale uncertainties as well as variations over
    switch-off functions. }
  \label{fig:threshold_full}  
\end{figure}
pair, fixed-order perturbation theory is not a good
approximation. Very close to threshold, the effective field theory of
(v/p)NRQCD separates the hard scale $m_t$, the soft scale given by the
top momentum of the non-relativistic top quark with velocity $v$, $m_t
v$ and the ultrasoft scale, given by the kinetic energy of the top
quark, $m_t v^2$ and allows to resum large logarithms of $v$
with $\alpha_s \sim v \sim 0.1$ close to
threshold. "Fixed-order" calculations resumming only Coulomb
singularities, but no velocity logarithms, for the totally inclusive 
$t\bar{t}$ production have been carried out in NRQCD to
NNNLO~\cite{Beneke:2015kwa}. The large velocity logarithms have been
resummed to next-to-next-leading logarithmic
(NNLL)~\cite{Hoang:2013uda} order
(cf. also~\cite{Hoang:2001mm,Pineda:2006ri} for predictions which did
not contain the full set of NNLL ultrasoft logarithms). These NRQCD 
calculations are based on the optical theorem and hold only for the
total inclusive cross section and in a narrow window around the
$t\bar{t}$ threshold. We report here about work where we combined and
matched the NLL NRQCD-resummed process close to the top threshold with
the fixed-order (relativistic) QCD-NLO process in the continuum. By a
carefully performed matching procedure, our approach smoothly
interpolates between the threshold region and the continuum, and it
allows to study all kinds of differential distributions. 

The matching is embedded into the \whizard-\openloops\ QCD-NLO
fixed-order framework discussed in Sec.~\ref{sec:continuum}. The NLL
resummed NRQCD contributions are included in terms of (S-/P-wave) form
factors to the (vector/axial vector) $\gamma/Z-t-\bar{t}$ vertex. These
form factors are obtained from the numerical solution of 
Schr\"odinger-type equations for the NLL Green functions computed
by the 
\texttt{Toppik}~\cite{Jezabek:1992np,Harlander:1994ac,Hoang:1999zc}
code, which is included in \whizard. The technical details of their
implementation and the matching setup can be found
in~\cite{Bach:2017ggt}. In order to avoid double-counting between
the fixed-order QCD-NLO part and the resummed NLL-NRQCD part, one has
to expand the form factors to first order in $\alpha_s$ and subtract
those pieces. As the NRQCD resummed calculations are not available for
the 5-point functions $\gamma^*/Z^* \to W^+ b W^- \bar{b}$, but only
for the top-vector and axial-vector currents, this removal of
double-counting has to be done in a factorized approach within a
double-pole approximation. There is no trivial gauge-invariant subset
for the process $e^+e^- \to W^+ b W^- \bar{b}$. In order to maintain
gauge-invariance of the factorized amplitudes, an on-shell projection
of the exclusive final states to the top mass shell is 
performed. The technical details, especially concerning the
direction of the three-momenta and the definition of the on-shell
projection below the kinematical threshold can be found
in~\cite{Bach:2017ggt}. The implementation inside \whizard\ has been
\begin{figure}
  \begin{center}
    \includegraphics[width=.48\textwidth]
                    {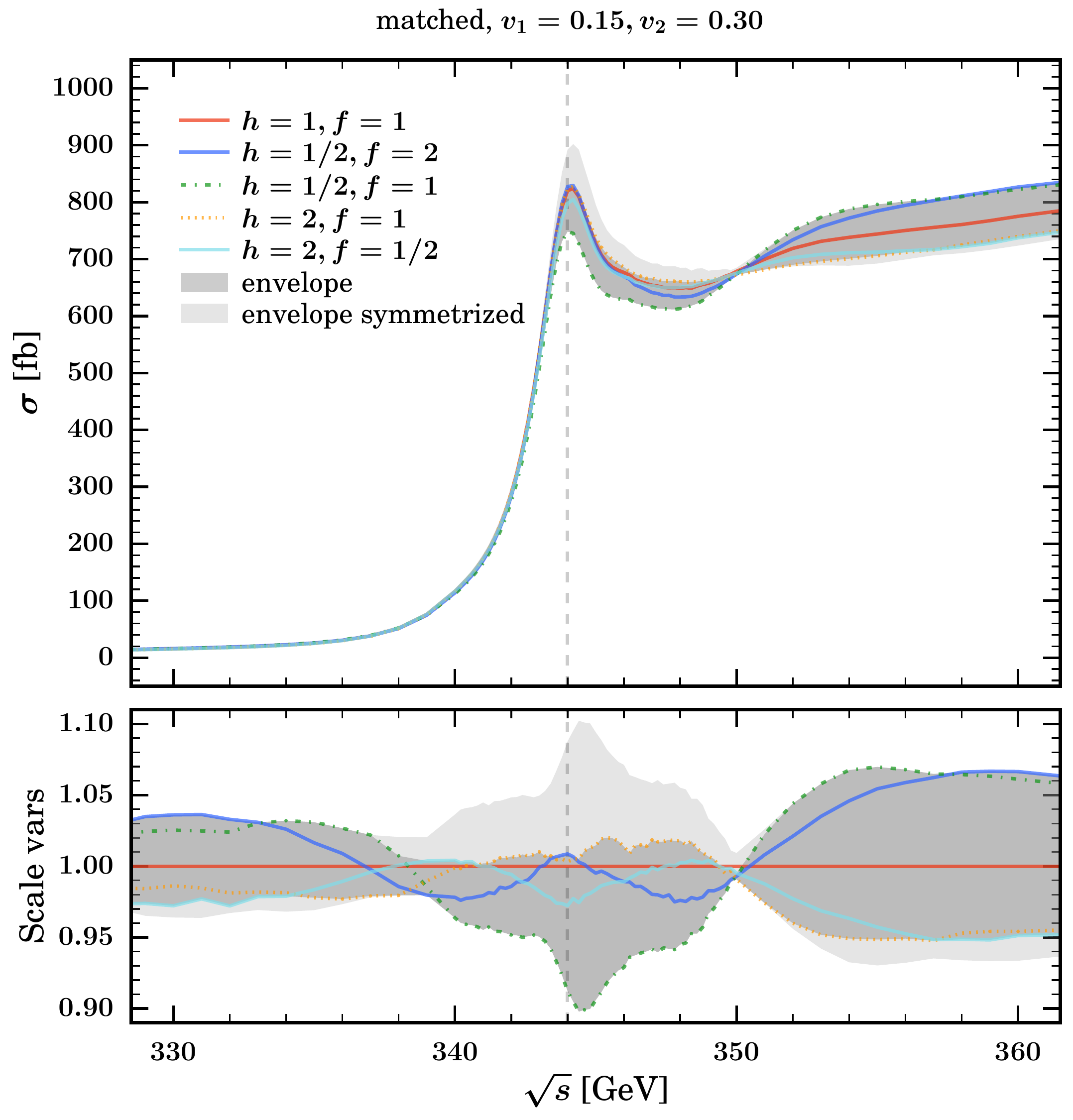}
    \includegraphics[width=.48\textwidth]
                    {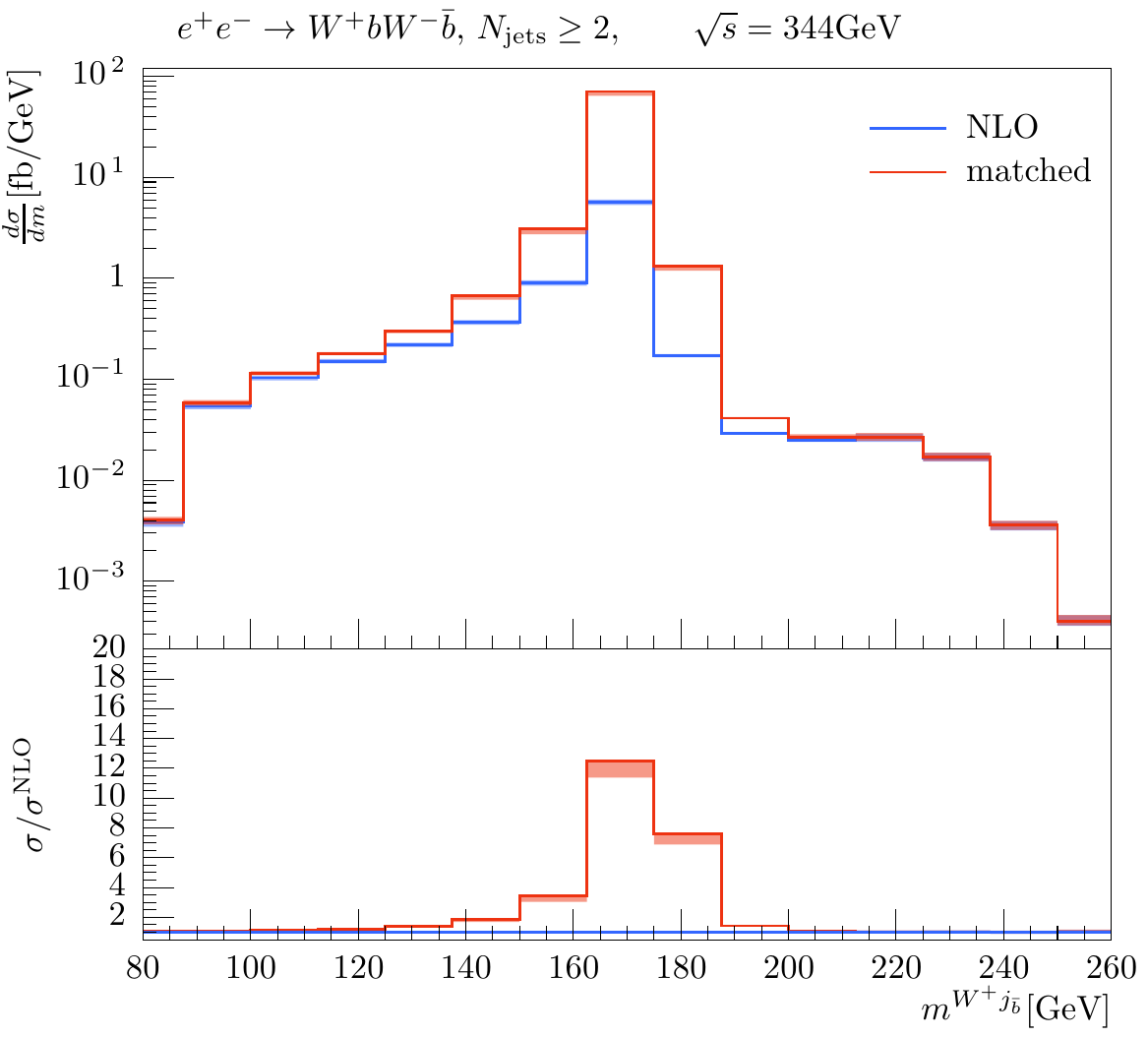}    
  \end{center}
  \caption{Left panel: Matched NRQCD-NLL + QCD-NLO total cross section
    as on the left of Fig.~\ref{fig:threshold_full}, but for a single
    choice of switch off-function. $h$ and $f$ are renormalization
    scale parameters as defined in
    \cite{Bach:2017ggt,Hoang:2013uda}. The grey bands display the
    corresponding scale 
    variations with and without symmetrization. Right panel: $Wb$
    invariant mass distribution at threshold ($\sqrt{s}=344$ GeV) as
    obtained with \whizard. The red line represents the full NRQCD-NLL
    + QCD-NLO matched, and the blue line the pure QCD-NLO result. The
    associated bands are generated by the same scale variations as in
    the left panel, here without symmetrization.}
  \label{fig:match_diff}
\end{figure}
validated with analytical calculations for different invariant mass
cuts on the reconstructed top quarks from Ref.~\cite{Hoang:2010gu}.

As for larger values of the top velocity ($v \gtrsim 0.4$) 
only the relativistic QCD-NLO result is valid, we define a switch-off
function that smoothly interpolates between the two regions. This
function is in principle arbitrary, and the possibility to vary this
function and its parameters adds another theory uncertainty on top of
the different scale variations in the different kinematic regimes. For
technical details again, we refer to~\cite{Bach:2017ggt}. The
results of our matching procedure are displayed in
Fig.~\ref{fig:threshold_full}. These plots 
show the total inclusive cross section for the process $e^+e^- \to W^+ b W^-
\bar{b}$, in the left panel without and in the right panel with QED
initial-state radiation (ISR). The dashed vertical line gives the
value for $2 M^{1\text{S}}$. The 1S mass $M^{1\text{S}}$ is defined as
half of the perturbative mass of a would-be 1S toponium state and
represents a renormalon free short-distance mass, which we treat as an
input parameter in \whizard. The blue
line shows the QCD-NLO cross section including scale variations in the
blue shaded areas. The only difference to the results in
Sec.~\ref{sec:continuum} is a different renormalization,
see~\cite{Bach:2017ggt}. The red curve shows the NRQCD-NLL + QCD-NLO
result, while the shaded band contains all (symmetrized) scale
variations of the hard, soft and ultrasoft
factorization/renormalization scales according
to~\cite{Hoang:2013uda} as well as variations of the switch-off
function to a reasonable extent~\cite{Bach:2017ggt}. The dotted black
line shows the matched results without applying a switch-off function
to the factorized NRQCD terms which deviates above threshold from
the relativistic QCD-NLO result. In Fig.~\ref{fig:match_diff}, left
panel, we see the matched result in the threshold region for a
single choice  of switch-off parameters, but scale variations over the
full two-dimensional renormalization parameter range defined
in~\cite{Hoang:2013uda}. This shows that the 
scale variation bands for the resummed NLL result in the threshold
region are highly asymmetric with respect to the central value which
motivates to apply a symmetrization of the error bands around the
central value. This symmetrization is also shown in
Fig.~\ref{fig:threshold_full}. In the right panel of
Fig.~\ref{fig:match_diff} we show as an example for a differential
distribution the invariant mass of the $W-b$ jet system. Blue is the
fixed-order QCD-NLO distribution, while red is the fully matched
distribution including scale variations, here un-symmetrized. The
ratio plot in the bottom does not show a K factor, but the ratio of
the matched result to the QCD-NLO fixed order result. It shows an
enhancement in the top mass peak due to threshold resummation by a
factor of ten to twelve.

 
\section{Summary and Outlook}
\label{sec:summary}

We presented work on the QCD-NLO corrections for exclusive top-quark
pair production including top and (leptonic) $W$ decays. Kinematic
regions in the continuum up to the highest available energies of CLIC
were covered as well as the threshold region. Both calculations have
been performed in the QCD-NLO framework of the \whizard\ event
generator which allows to include all important physics of a lepton
collider like polarization, QED ISR radiation and non-trivial beam
spectra. The continuum calculations represent the first massive $2\to
6$ and $2\to 7$ QCD-NLO calculations for lepton colliders. The matched
threshold calculation smoothly interpolates the threshold region
described by non-relativistic QCD to the relativistic QCD-NLO
calculation and constitutes the highest available precision available
at the level of the completely exclusive final state. While the work
presented here is more a proof-of-principle of the matching procedure
between threshold and continuum, it is obvious that the various
differential distributions (which are accessible now and of which we
only showed one here) offer plenty of possibilities for top mass
measurements. This is part of ongoing and future work. Other
directions for future work are the matching at even higher order in
QCD, the inclusion of electroweak corrections to the relativistic
continuum process, and the inclusion of $W$ decays also in the matched
calculations.


\section*{Acknowledgments}

JRR wants to thank the organizers of LCWS 2017 in Strasbourg for a
great conference at a lovely venue. JRR, BCN and CW acknowledge
support from the Collaborative Research Unit (SFB) 676 of the DFG,
projects B1 and B11. We like to thank Stefan Kallweit for his help
with the cross-checks by the \texttt{Munich} code.


\end{document}